\documentstyle[12pt,epsfig]{article}

\newlength{\dinwidth}
\newlength{\dinmargin}
\def\lapproxeq{\lower .7ex\hbox{$\;\stackrel{\textstyle
<}{\sim}\;$}}
\def\gapproxeq{\lower .7ex\hbox{$\;\stackrel{\textstyle
>}{\sim}\;$}}

\begin{document}
\titlepage
\begin{flushright}
DTP/96/40  \\
May 1996 \\
\end{flushright}
\vspace*{2cm}

\begin{center}
{\Large\bf The Partonic Structure of the Proton \footnotemark}\\
\vspace*{1cm}
A.\ D.\ Martin\\
\vspace*{0.5cm}
Department of Physics, University of Durham,\\
Durham, DH1 3LE, England\\
\end{center}

\footnotetext{To be published in the Proceedings of 
the 2$^{nd}$ Krak\'ow Epiphany Conference on \lq\lq 
Proton Structure", January 1996 in Acta Physica Polonica}

\vspace*{2cm}
\begin{abstract}
We review the latest information that is available about the parton 
distributions of the proton, paying particular attention to the 
determination of the gluon. We briefly describe the various processes
that have been advocated to be a measure of the gluon. We discuss the importance 
of the gluon to the description of the structure function $F_2$ at small
$x$, with emphasis on the $\ln 1/x$ resummations.
\end{abstract}

\newpage
\begin{center}
\section{Parton distributions}
\end{center}

Perturbative QCD is remarkably successful in describing the broad 
sweep of hard scattering processes involving the proton. A vital common 
ingredient is a universal set of parton distributions, $f_i (x, Q^2)$,
which allow all of these reactions to be calculated in terms of basic 
QCD subprocesses at the partonic level. $f_i (x, Q^2)$ is the 
probability of finding parton $i$ (where $i$ may be a quark, antiquark 
or gluon) within the proton carrying a fraction $x$ of its momentum when
probed by a particle with virtuality $Q^2$.

The classic way to probe the partonic structure of the proton is 
deep-inelastic lepton-nucleon scattering, where the lepton may be an 
electron, a muon or a neutrino. At high energy the differential cross-section
for, say, deep-inelastic electron-proton scattering ($e p \rightarrow e X$)
has the form
\begin{equation}
\frac{d^2\sigma}{dxdQ^2} = \frac{4\pi\alpha^2}{xQ^4} \left [ y^2\; xF_1(x, Q^2)
+ (1 - y)\; F_2 (x, Q^2) \right ]
\end{equation}
with $Q^2 \equiv -q^2$, the Bjorken $x$ variable $x = Q^2 / 2p.q$ and 
$y = Q^2 / xs$,
where $p$ and $q$ are the 4-momenta of the proton and virtual exchanged 
photon respectively. $\sqrt{s}$ is the centre-of-mass energy of the 
electron-proton collision. It is easy to see that the momentum fraction $x$ is
the same as the Bjorken $x$. Since the struck quark acquires 4-momentum 
$xp + q$ we have $(xp + q)^2 = m_q^2$. Thus $x = Q^2 / 2p.q$ in the infinite
momentum frame where masses may be disregarded.

The relation between the observable structure function $F_2$ and the parton 
densities $f_i$ is, to $O(\alpha_S)$, of the form
\begin{equation}
F_2(x, Q^2) = x \sum_q e_q^2 \int_x^1 \frac{d\xi}{\xi}\; \left\lbrace f_q (\xi, 
Q^2)\;
\left [ \delta \left (1 - \frac{x}{\xi}\right )\; + \frac{\alpha_S}{2\pi}\; C_q 
\left (\frac{x}{\xi}\right ) \right ] + f_g \left ( \xi, Q^2\right )\;
\frac{\alpha_S}{2\pi}\; C_g \left (\frac{x}{\xi}\right ) \right\rbrace,
\end{equation}
\begin{figure}[htb] 
\begin{center}
~ \epsfig{file=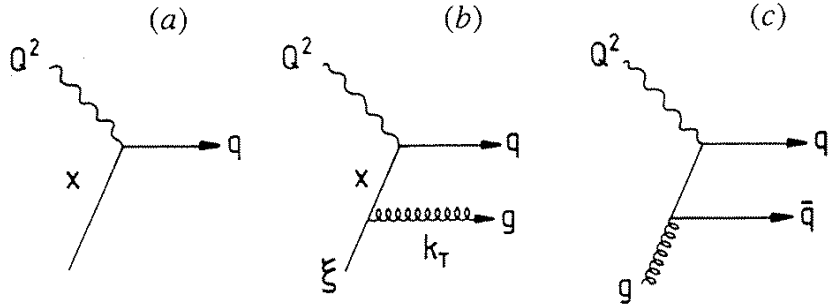,width=12cm}
\end{center}
\caption{\protect{\small Partonic subprocesses which contribute to 
deep-inelastic 
scattering: (a) the lowest-order diagram, which is responsible for the quark 
parton model and (b), (c) QCD diagrams of first order in $\alpha_S$ that give 
contributions to $F_2$ which depend on the quark and gluon content 
respectively of the proton.}}
\label{fig1}
\end{figure}
\noindent where the partonic subprocesses are shown in Fig. 1.  The $O 
(\alpha_S)$ QCD 
subprocesses shown in (b,c) have initial state 
collinear singularities, which are factored off into the parton densities 
causing them to \lq\lq run" (i.e. to depend on $Q^2$) leaving
well-behaved known coefficient functions $C_q$ and $C_g$. Due to this 
renormalisation, the absolute values of the parton densities are not calculable
in perturbative QCD. Rather QCD determines the $Q^2$ dependence (or
so-called scaling violations). It is given by the DGLAP evolution equations 
\cite{DGLAP} which have the form
\begin{eqnarray}
\frac{\partial f_i (x, Q^2)}{\partial \ln Q^2} &=& \sum_j \int 
\frac{dx'}{x'}\; P_{ij} \left(\frac{x}{x'}\right )\; 
f_j (x', Q^2)\nonumber \\
&\equiv& P_{ij} \otimes f_j,
\end{eqnarray}
where the splitting functions
\begin{equation}
P_{ij} = \alpha_S\; P_{ij}^{(1)}\; + \; \alpha_S^2\; P_{ij}^{(2)}\; + ...
\end{equation}
So far, the leading order (LO), $P_{ij}^{(1)}$, and next-to-leading order (NLO),
$P_{ij}^{(2)}$, terms have been calculated.

Effectively, the $P_{ij}^{(1)}$ term resums the leading $\ln Q^2$ terms. 
That is the $(\alpha_S \ln Q^2)^n$ contributions which, in an axial gauge, 
correspond to the sum of ladder diagrams (with $n$ rungs) in which the 
transverse momenta of the emitted partons (gluons) are strongly ordered along 
the chain (i.e. $Q^2 \gg k_{nT}^2 \gg ... \gg k_{1T}^2$ in the example shown in 
Fig. 2). The NLO contribution corresponds to the case when a pair of momenta 
are comparable $k_{iT} \approx k_{i+1T}$ and we lose a power of $\ln Q^2$.
That is $P_{ij}^{(2)}$ sums up the $\alpha_S^n \ln^{n-1} Q^2$ contributions. 
When truncating the power series at a given power of $\alpha_S$, 
say $\alpha_S^m$, the renormalisation of the $f_i$ introduces a scheme 
dependence of $O(\alpha_S^{m+1})$. Traditionally the $\overline{\rm MS}$
scheme is used.

\begin{figure}[htb] 
\begin{center}
~ \epsfig{file=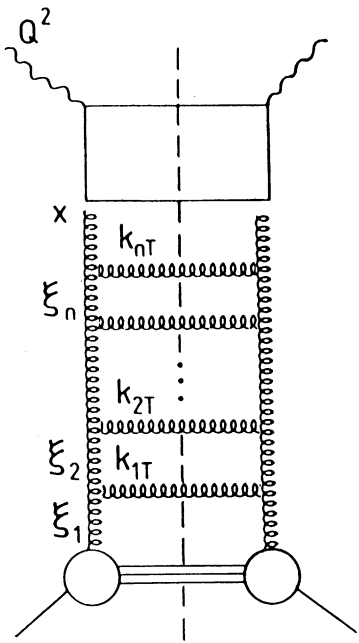,height=9cm}
\end{center}
\caption{\protect{\small DGLAP evolution, which at LO sums the 
$(\alpha_S \ln Q^2)^n$ 
contributions, corresponds for $P_{gg}$ to the sum of such ladder diagrams 
with the transverse momenta of the emitted gluons strongly ordered along 
the chain $(Q^2 \gg k_{nT}^2 \gg ... \gg k_{1T}^2)$. On the other hand for 
the BFKL $\ln 1/x$ summation (which is discussed in section 4) the ladder 
must be regarded as an effective ladder diagram incorporating
many different contributions. In this case the $(\alpha_S \ln 1/x)^n$
contribution comes from the strongly ordered configuration 
$x \ll \xi_n \ll ... \ll \xi_1$ but with the gluon $k_T$ values unordered.}}
\label{fig2}
\end{figure}

\begin{center}
\section{Global analyses}
\end{center}

The parton densities describe not only deep-inelastic scattering, but all 
hard scattering processes with incoming nucleons. As we have noted, the
densities $f_i (x,Q^2)$ have to be determined by experiment at some scale
$Q^2 = Q^2_0$. The basic procedure is to parametrize the $x$ dependence at
some low $Q^2_0$, but where perturbative QCD should be applicable, and then
to evolve up in $Q^2$ using the NLO DGLAP equations to determine $f_i (x,Q^2)$ 
at all the values of $x, Q^2$ of the data. The input parameters are then 
determined by a global fit to the data.

To be specific the 1994/5 MRS \cite{MRSA,MRSG} and CTEQ \cite{CTEQ} analyses 
took $Q^2_0$ = 4 GeV$^2$ for the input scale. We describe the MRS analyses.
Similar results are obtained by CTEQ. The starting distributions are taken 
to be of the form
\begin{equation}
x\; f_i(x,Q^2_0) = A_i\; x^{-\lambda_i} (1 - x)^{\beta_i}\; (1 + \gamma_i 
\sqrt{x} + \delta_i\; x)
\end{equation}
for $i = g$, $u_{\rm val}$, $d_{\rm val}$ and the (total) quark sea $S$. In
practice not all of the parameters $A_i,\;\lambda_i,\;\beta_i,\;\gamma_i, 
\;\delta_i$ are free. Three of the $A_i$ are determined by the flavour and
momentum sum rules. Moreover we have some idea of the values of the
$\beta_i$ and $\lambda_i$ from spectator counting rules and Regge
expectations respectively. The QCD coupling is also a free parameter. It is
determined primarily by the scaling violations observed in the high precision
BCDMS $F_2^{\mu p, \mu d}$ data in the region $0.35 \lapproxeq x \lapproxeq 
0.55$.

The flavour structure of the quark sea $S = 2(\bar{u} + \bar{d} + \bar{s} + 
\bar{c} + ...)$ is determined by data. The CCFR dimuon production data 
\cite{CCFR} imply that the strange sea is suppressed by 0.5 relative to the
$u$ and the $d$ sea distributions at $Q^2$ = 4 GeV$^2$. The difference 
$\bar{d} - \bar{u}$ is arranged to be compatible with the observed NA51
\cite{NA51} asymmetry in Drell-Yan production in $pp$ and $pn$ collisions.
The input charm sea is determined by EMC deep-inelastic data for
$F_2^c$ \cite{EMC}. We assume $c = 0$ for $Q^2 < m^2$ and for higher $Q^2$ 
we generate $c(x,Q^2)$ by massless evolution. The data imply $m^2$ = 2.7
GeV$^2$. After evolution to $Q^2$ = 4 GeV$^2$ we find that the charm sea, 
to a good approximation, satisfies $2c$ = 0.02$S$. The description of the
EMC charm data by MRS(A) partons is shown in Fig. 3. Clearly this is an
approximate way to treat $m_c \neq 0$ effects. At this meeting De Roeck 
\cite{DER} presented the first preliminary measurements of $F_2^c$ by the
H1 collaboration. To get some idea of the future impact of these data,
estimates of the preliminary measurements of $F_2^c$ at $Q^2$ = 13, 23 and 
50 GeV$^2$ have been superimposed on the $x$ = 0.002 curve in Fig. 3. Although 
the present parton treatment of charm appears to be satisfactory, it is clear 
that future more precise data will be invaluable in the investigations of the 
proper treatment of $m_c \neq 0$ effects. In summary, the data imply that at 
the 
input scale, $Q^2_0$ = 4 GeV$^2$, the charm sea carries about 0.4\% of the
proton's momentum, as compared to nearly 4\% by the strange sea, 6\% by
the up sea and 9\% by the down sea.

\begin{figure}[htb] 
\begin{center}
~ \epsfig{file=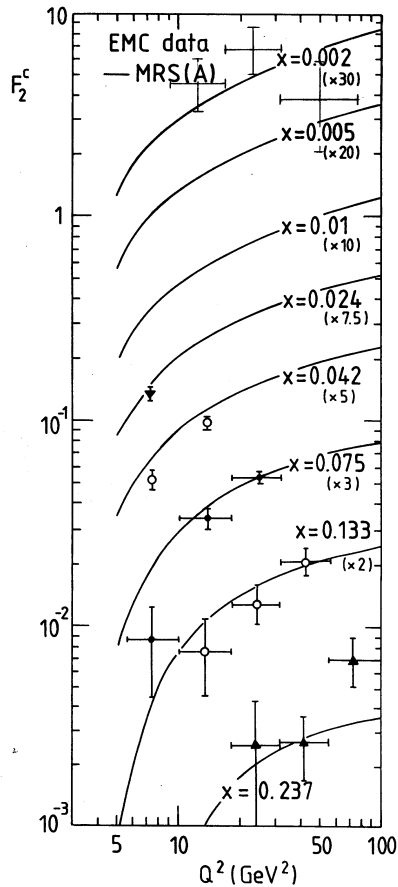,bbllx=0pt,bblly=0pt,bburx=300pt,bbury=750pt,%
height=12cm}
\end{center}
\caption{\protect{\small The description of the EMC measurements 
\protect{\cite{EMC}} of $F_2^c$ with $x \geq$ 0.024
by the MRS(A) partons \protect{\cite{MRSA}}. The preliminary measurements
of $F_2^c$ by the H1 collaboration, presented for the first time at this 
Conference \protect{\cite{DER}}, have been used to superimpose three
data points to be compared with the $x$ = 0.002 curve.}}
\label{fig3}
\end{figure}

\begin{table}[p]
\caption{\protect{\small The experimental data (in the top part of table) 
used to determine parton distributions in the global analyses. The last column 
gives an indication of the main type of constraint imposed by a particular set 
of data.  The processes in the bottom part of the table are discussed in 
Section 3.}}
\vspace{\bigskipamount}
\centering
\def\btop{\mathchar"1339}
\def\bmid{\mathchar"133D}
\def\bbot{\mathchar"133B}
\def\bs{\mathchar"1343}
\begin{tabular}{lll}    \hline
                      &Leading-order&                                   \\
Process and Experiment&subprocess   &Parton and $\alpha_S$ determination\\
\hline
&\hfill \raisebox{-0.5ex}[0.5ex][0.5ex]{$\btop$}&\\
{\bf DIS} $\mbox{\boldmath $(\mu N \rightarrow \mu X)$}$ &  $\gamma^*q
\rightarrow q$ \hfill {\arrayrulewidth=1pt\vline}\hspace*{4pt}& Four structure
functions $\rightarrow$ \\
BCDMS, NMC, E665& \hfill {\arrayrulewidth=1pt\vline}\hspace*{4pt}& 
\hspace*{0.5cm}
$u + \bar{u}$, $d + \bar{d}$ \\
$F^{\mu p}_2,F^{\mu n}_2$& \hfill $\bmid$ & 
\hspace*{0.5cm}  $\bar{u} + \bar{d}$, $s$ (assumed = $\bar{s}$) \\
{\bf DIS} $\mbox{\boldmath $(\nu N \rightarrow \mu X)$}$ & $W^*q \rightarrow
q^{\prime}$  \hfill {\arrayrulewidth=1pt\vline}\hspace*{4pt}& but only
$\int xg(x)dx \simeq 0.5$ \\
CCFR (CDHSW)&\hfill {\arrayrulewidth=1pt\vline}\hspace*{4pt}& 
[$\bar{u}-\bar{d}$ is not determined] \\
$F^{\nu N}_2,xF^{\nu N}_3$&\hfill \raisebox{0.5ex}[1.5ex][1.5ex]{$\bbot$}&
$\alpha_S$ ($x \approx$ 0.4 data) \\
$\mbox{\boldmath $\mu N \rightarrow c \overline{c} X$}$ &
$\gamma^*  c\rightarrow c$&$c \approx 0.1 s$  at $Q^2$ = 4 GeV$^2$\\
$F_2^{c}$, EMC&&\\
$\mbox{\boldmath $\nu N \rightarrow \mu^+\mu^-X$}$ &  $W^* s \rightarrow c$& 
$s \approx \frac{1}{2}\bar{u}$ (or $\frac{1}{2}\bar{d}$) \\
CCFR&$\;\;\;\;\;\;\;\;\;\;\;\;\;\hookrightarrow \mu^+$&\\
{\bf DIS (HERA})&$\gamma^*q \rightarrow q$&$\lambda_g$, $\lambda_S$, 
$\alpha_S$\\
$F^{ep}_2$ (H1,ZEUS)&&($xg \sim x^{-\lambda_g}, x\bar{q}\sim x^{-\lambda_S}$)\\
$\mbox{\boldmath $p p \rightarrow \gamma X$}$\hspace*{0.5cm}WA70 (UA6,&  
$qg \rightarrow \gamma q$ &$g(x \approx 0.4)$ \\
E706, R806, UA2, CDF)&&\\
$\mbox{\boldmath $pN \rightarrow \mu^+\mu^- X$}$   &  $q\bar{q} \rightarrow
\gamma^*$  &  $\bar{q} = ...(1-x)^{\beta_S}$ \\
E605&&\\
$\mbox{\boldmath $pp, pn \rightarrow \mu^+\mu^- X$}$ & $u\bar{u},d\bar{d}
\rightarrow \gamma^*$   & $(\bar{u}-\bar{d})$ at $x = 0.18$  \\
NA51&$u\bar{d},d\bar{u} \rightarrow \gamma^*$& \\
$\mbox{\boldmath $p\overline{p} \rightarrow W^{\pm}$}$ {\bf asym}&
$u\bar{d} \rightarrow W^+$&slope of $u/d$ at $x \approx 0.05$  \\
CDF&$d\bar{u} \rightarrow W^-$&\\
\hline
&&\\
$\mbox{\boldmath $p\overline{p} \rightarrow {\rm\bf jets}$}$&
$gg$, $gq$, $q\bar{q}$&
$g(0.005 \lapproxeq x \lapproxeq 0.1)$\\
CDF, D0&$\rightarrow$ 2 jets&$q(x \sim 0.2)$\\
&&$\alpha_S(E_T \sim 100$ GeV)\\
$\mbox{\boldmath $\gamma^* p \rightarrow {\rm\bf dijets}$}$&
$\gamma^* g \rightarrow q \bar{q}$&
$g(0.005 \lapproxeq x \lapproxeq 0.1)$, $\alpha_S$\\
H1, ZEUS&$\gamma^* q \rightarrow g \bar{q}$&\\
$\mbox{\boldmath $\gamma^* p \rightarrow {\rm J/}\psi X$}$&
$\gamma^* g \rightarrow (c \bar{c}) g$&$g$(?)\\
EMC, HERA&&\\
$\mbox{\boldmath $\gamma p \rightarrow {\rm J/}\psi p$}$&
$c \bar{c} \rightarrow c \bar{c}$&$g(x \sim 10^{-3})$\\
H1, ZEUS&via $gg$ exch.&\\
\hline
\end{tabular}
\end{table}

The wide range of data used in the global fits is shown in the top 
part of Table 1, together with an indication of the most important constraints
that they impose on particular partons. Fig. 4 shows the parton distributions 
at
$Q^2$ = 20 GeV$^2$ corresponding to the 1994 and 1995 sets of MRS
partons \cite{MRSA,MRSG}, which were obtained from global fits to these data.
\begin{figure}[ht] 
\begin{center}
~\epsfig{file=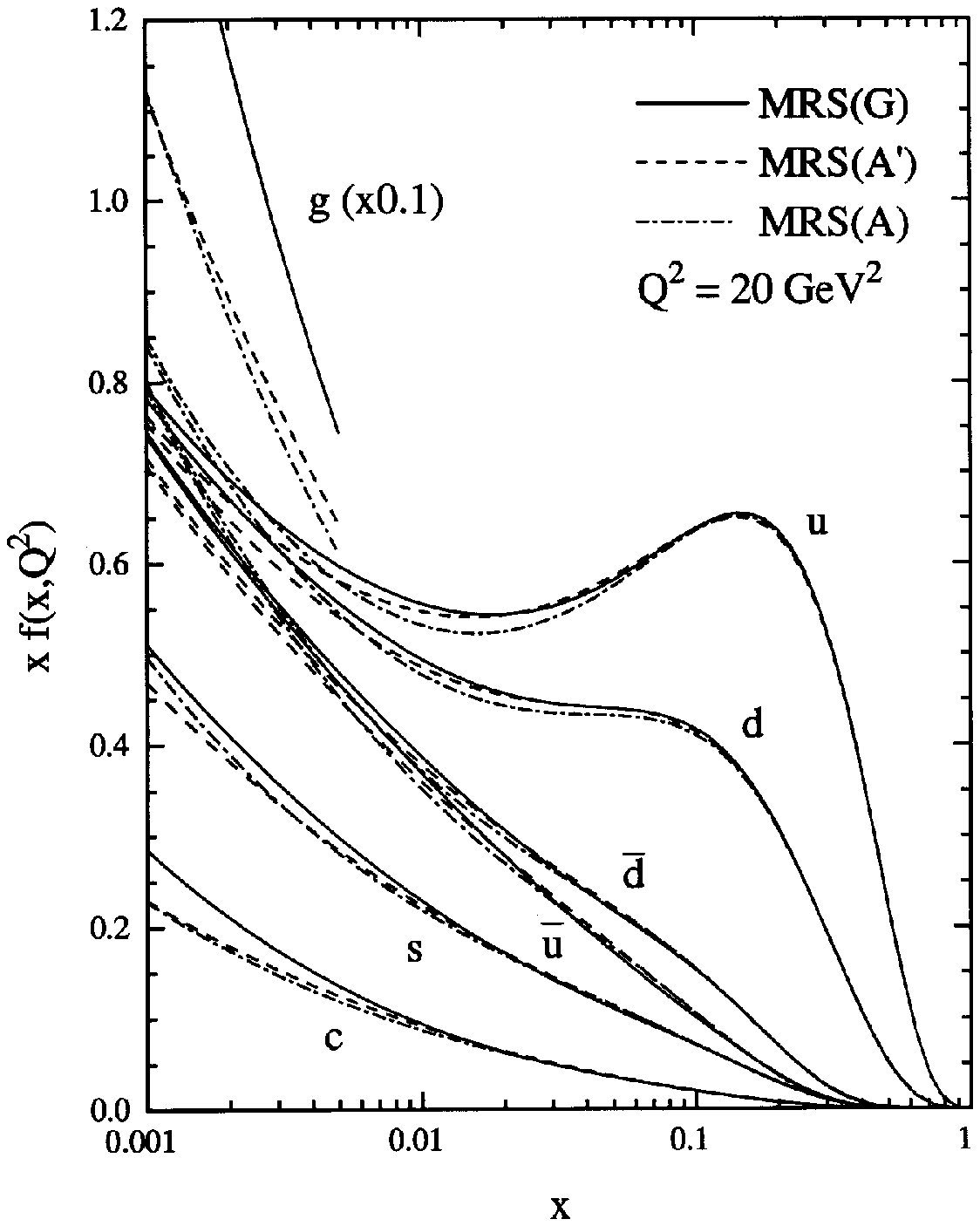,bbllx=0pt,bblly=0pt,bburx=360pt,bbury=425pt,%
height=12cm}
\end{center}
\caption{\protect{\small The 1994 \protect{\cite{MRSA}} and 1995 
\protect{\cite{MRSG}} sets
of MRS partons at $Q^2$ = 20 GeV$^2$. For clarity the gluon distribution
(divided by a factor of 10) is only shown for $x <$ 0.005.}}
\label{fig4}
\end{figure}
The differences $u - \bar{u} \equiv u_{\rm val}$ and $d - \bar{d} \equiv 
d_{\rm val}$ show 
the valence quark structures around $x \sim$ 0.1. The dominance of the 
gluon for $x \leq$ 0.01 is also evident.  Note that the gluon is suppressed 
on the figure by a factor of 10.  The general conclusion is that the partons 
are well determined for $x \gapproxeq 0.02$, where data for a wide range of 
processes exist, except possibly the gluon which in this region is mainly 
constrained by prompt photon data. The gluon is clearly the crucial parton in 
the small $x$ domain. It is the subject of the next two sections.

\begin{center}
\section{Determination of the gluon}
\end{center}

The gluon only contributes at leading order in prompt photon production
among all the processes fitted in the global analyses, see Table 1. Its
distribution is therefore not so well determined as those of the quarks.
The constraints on the gluon come mainly from (i) the momentum sum
rule, (ii) prompt photon production, and (iii) the scaling violations of
$F_2$. Scaling violations impose the tightest constraint in 
the small $x$ region, where the gluon is the dominant parton. Then
\begin{equation}
\frac{\partial F_2}{\partial\ln Q^2} \approx P_{qg} \otimes g
\end{equation}
where the convolution leads to the gluon being sampled at a higher value
of $x$ than that at which the violation is measured. Roughly speaking 
an observed violation at $x$ measures $\alpha_S (Q^2)$ $g (2x)$. In this
way the HERA measurements of the scaling violations of $F_2$ at small $x$
have considerably improved our knowledge of the gluon.

\begin{figure}[t] 
\begin{center}
~\epsfig{file=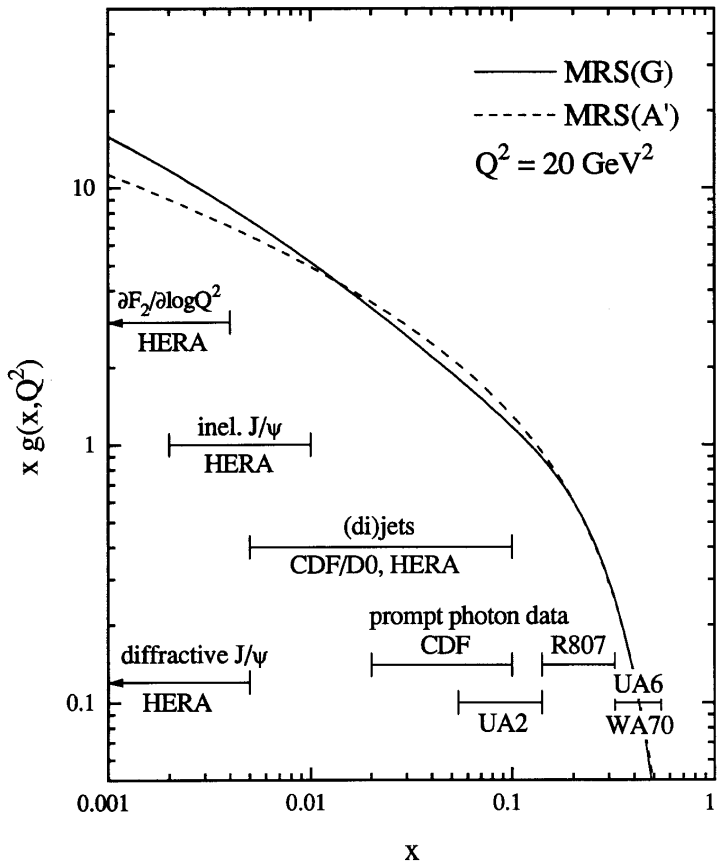,bbllx=0pt,bblly=30pt,bburx=360pt,bbury=440pt,%
height=12cm}
\end{center}
\caption{\protect{\small The $x$ intervals in which the gluon may be 
constrained by various sets of data. Also shown are the gluons at 
20 GeV$^2$ from ref.~\protect{\cite{MRSG}}.}}
\label{fig5}
\end{figure}

These and other potential determinations of the gluon are summarised in Fig. 5,
with an indication of the relevant $x$ ranges. We discuss the determinations
in turn below.

\begin{center}
\subsection{$F_2$ scaling violations at HERA}
\end{center}

In the early NLO global fits not all the parameters in
\begin{equation}
xg = A_g x^{-\lambda_g}\; (1 - x)^{\beta g}\; (1+ \gamma_g \sqrt{x} + 
\delta_g x)
\end{equation}
were used.  In particular, the data did not determine the small $x$ 
behaviour of the gluon. For example, $\gamma_g$ was set to zero, and 
since in the perturbative region the gluon drives the sea, via 
$g \rightarrow q \bar{q}$, it was assumed that $\lambda_g = \lambda_S$ at
the input scale $Q^2_0$. With the advent of the HERA measurements of $F_2$,
and their improvement year-by-year, the gluon has become better and 
better determined in the small $x$ region. 
The improvement is reflected in Table 2 in the step-by-step release of the
parameters, $\lambda_g$ and $\gamma_g$, which most affect the small $x$
behaviour of the gluon.  The G set of partons allowed $\lambda_S \neq \lambda_g$ for the first time,
but it lead to only a marginal improvement with respect to the set A$'$ in
which $\lambda_S$ was set equal $\lambda_g$ \cite{MRSG}. The new HERA data
\cite{H1,ZEUS} exclude the G set of partons and it is interesting to note
that the new parton set R1 \cite{MRSR} is similar to the A$'$ set of partons.

\begin{table}[htb]
\caption{\protect{\small The exponents $\lambda_i$ of $xg \sim x^{-\lambda_g}$ 
and $xS \sim x^{-\lambda_S}$ at the input scale $Q^2_0$ = 4 GeV$^2$ for a
sequence of MRS analyses \protect{\cite{MRSD,MRSA,MRSG,MRSR}} which include
more and more precise HERA data as they become available each year. The
latest fit \protect{\cite{MRSR}} has input scale $Q^2_0$ = 1 GeV$^2$ but the
exponents are also given after evolution up to $Q^2$ = 4 GeV$^2$. The values of
the parameters $\lambda_i$ are strongly correlated with the values of $\gamma_i$
(see the discussion in ref. \protect{\cite{MRSA}}).}}
\vspace{\bigskipamount}
\centering
\begin{tabular}{@{}llcccl}\hline
\multicolumn{2}{c}{MRS fit}&$\lambda_g$&&$\lambda_S$&\\
\hline
1993&D$_0$&0&=&0&fixed\hfill ($\gamma_g = 0$)\\
    &D$_-$&0.5&=&0.5&fixed\hfill ($\gamma_g = 0$)\\
    \\
1994&A    &0.3&=&0.3&free\hfill ($\gamma_g = 0$)\\
\\
1995&A$'$ &0.17&=&0.17&free\\
    &G    &0.31&&0.07&free\\
    \\
1996&R1   &(0.17)&&(0.18)&at $Q^2$ = 4 GeV$^2$\\
    &     &$-$0.55&&0.12&at $Q^2_0$ = 1 GeV$^2$\\
\hline
\end{tabular}
\end{table}

Traditionally the MRS and CTEQ analyses have fitted to data with $Q^2 > 5$ 
GeV$^2$.  In the GRV approach \cite{GRV,REYA} valence-like forms of the parton 
distributions are taken at a low input scale $Q_0^2 = 0.34$ GeV$^2$.  The 
original hope of this \lq\lq dynamical" model was that input valence quarks 
would suffice and that the gluon and sea distributions would be generated 
radiatively.  However, a sizeable valence gluon and a valence sea distribution 
are also required at the input scale in order to describe prompt photon and 
NMC deep-inelastic data respectively, which as a consequence introduces more 
phenomenological parameters into the GRV model.  The GRV partons were found to 
give a good 
description of the HERA data down to unexpectedly low values of $Q^2$, namely 
$Q^2 \sim 1.5$ GeV$^2$, although with the precision of the latest data there is 
some discrepancy at the lowest values of $x$, see Fig.\ 6.  Motivated by the 
general success of the GRV (DGLAP-based) predictions, 
the latest MRS analysis \cite{MRSR} uses a lower input scale, $Q_0^2 = 1$ 
GeV$^2$, 
and fits to data $Q^2 \geq 1.5$ GeV$^2$ --- the resulting description of the 
new HERA data at the lowest values of $x$ is shown in Fig.\ 6.  The continuous 
and dashed curves correspond to setting $\alpha_S (M_Z^2) = 0.113$ and 0.120 
respectively.  The first choice of $\alpha_S$ is the value determined by the 
scaling violations of the fixed-target deep-inelastic data, in particular the 
BCDMS $F_2^{\mu p, \mu d}$ measurements in the interval $0.35 \lapproxeq x 
\lapproxeq 0.55$ \cite{MV}.  The second choice is preferred by LEP data 
\cite{BET}, 
and also marginally by the HERA measurements of $F_2^{ep}$.  The two new MRS 
fits 
have both $\lambda_g$ and $\lambda_S$ as free parameters.  To see the extent 
to which they can be determined independently, fits are also performed with 
$\lambda_g = \lambda_S$.  The gluon distributions of these 4 fits at $Q^2 = 5$ 
GeV$^2$ are shown in Fig.\ 7.  Also shown in this plot is a representative 
spread of the gluons that were available in 1995.  The large difference 
between the A$^\prime$ (with $\lambda_S = \lambda_g$) and G (with $\lambda_S 
\neq \lambda_g$) gluons is not present in the new fits --- they all cluster 
about A$^\prime$.  The new HERA data appear to have significantly pinned 
down the gluon.

\begin{figure}[htb] 
\begin{center}
~ \epsfig{file=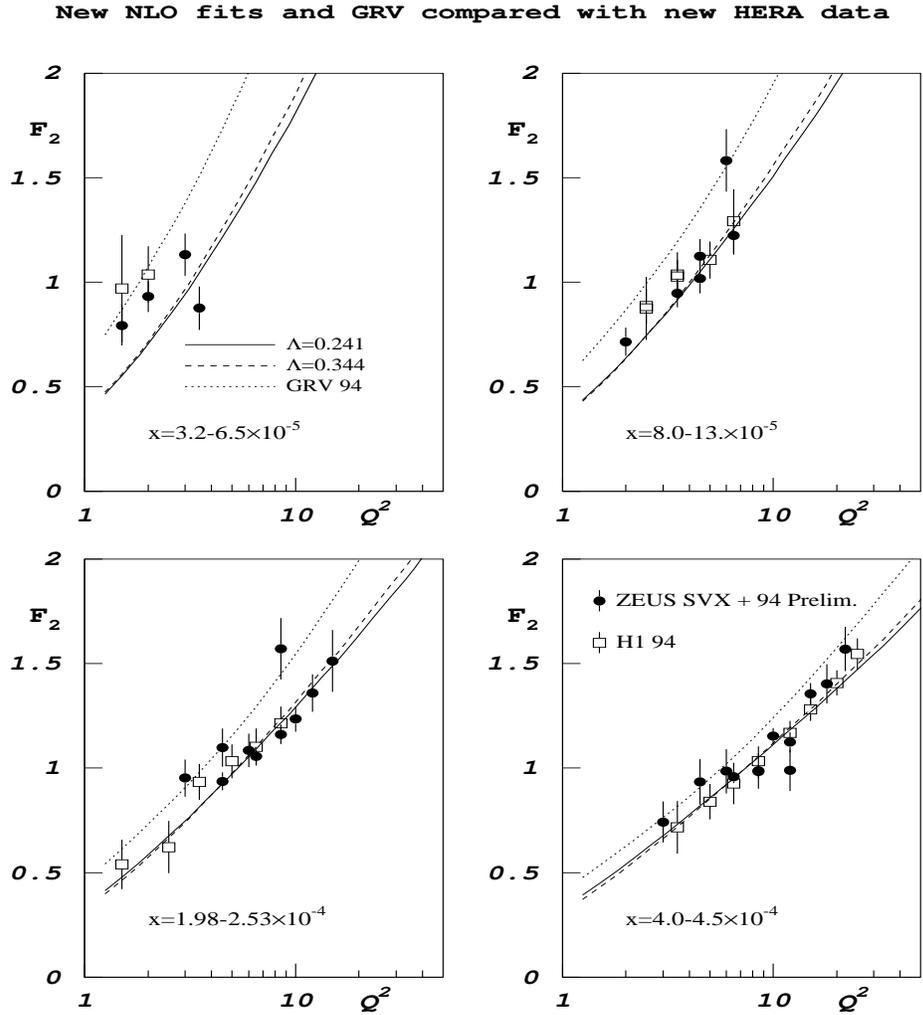,bbllx=0pt,bblly=90pt,bburx=500pt,bbury=750pt,%
height=14cm,width=14cm}
\end{center}
\caption{\protect{\small Recent HERA measurements \protect{\cite{H1,ZEUS}} of 
the proton structure function $F_2$ versus $\ln Q^2$ at the lowest values of $x$.  
The continuous and dashed curves are the description obtained in a new 
preliminary 
MRS global analysis with the QCD coupling taken to be such that 
$\alpha_S (M_Z^2) = 0.113$ and 0.120 respectively \protect{\cite{MRSR}}.  
The dotted curve is the prediction obtained from the GRV partons 
\protect{\cite{GRV}}.}}
\label{fig7}
\end{figure}

\begin{figure}[htb] 
\begin{center}
~ \epsfig{file=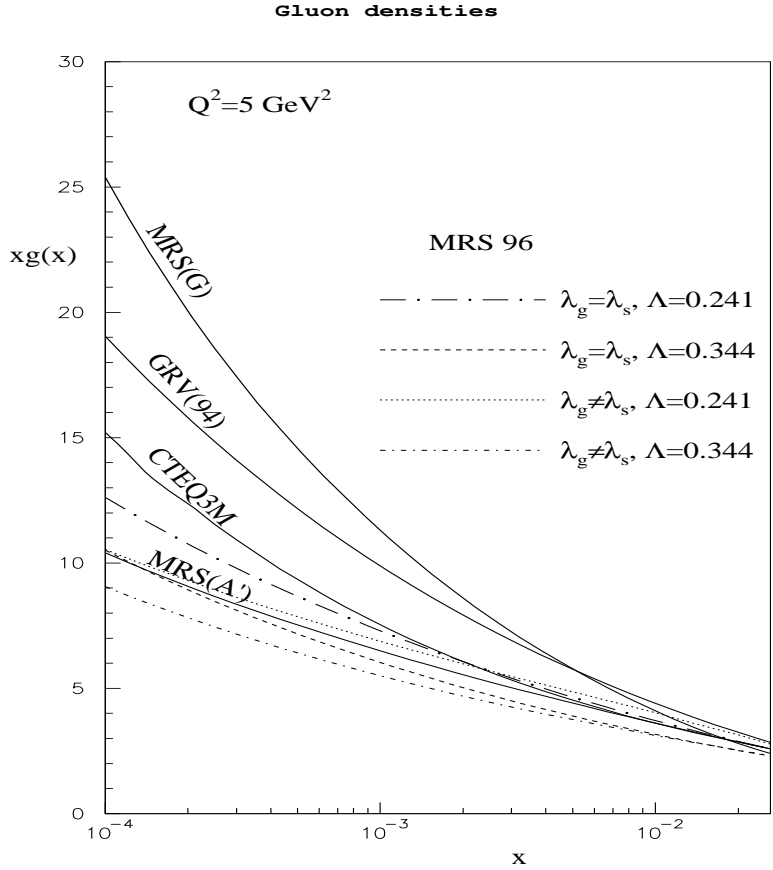,bbllx=0pt,bblly=70pt,bburx=500pt,bbury=750pt,%
height=12cm,width=12cm}
\end{center}
\caption{\protect{\small The gluon distribution at $Q^2 = 5$ GeV$^2$.  The four 
continuous curves show a representative set of the gluons that were available 
in 1995.  The four broken curves correspond to gluons obtained 
\protect{\cite{MRSR}} in global analyses which include the latest HERA data --- 
the new solutions cluster around the MRS(A$^\prime$) gluon.}}
\label{fig6}
\end{figure}

\begin{center}
\subsection{Prompt photon production}
\end{center}

The processes $pp \rightarrow \gamma X$ and $p\overline{p} \rightarrow \gamma X$ 
have long been regarded as a classic way to determine the gluon.  The 
perturbative 
QCD  formulae are now known to NLO, including the fragmentation (or 
Bremsstrahlung) 
contribution \cite{PPNLO}.  Moreover, the experiments (WA70, UA6, E706, R806, 
UA2, 
CDF) cover the entire $x$ interval from 0.6 down to 0.01.  The situation 
therefore 
appears promising.  However, there is a pattern of deviation in the shape of the $p_T$ 
dependence.  The data are steeper in $p_T$ than the QCD predictions.  Neither 
changes of scale nor the introduction of fragmentation effects can resolve the 
discrepancy\footnote{Vogelsang and Vogt \cite{VV} have demonstrated 
that these effects can improve the description of a single experiment.  However, 
experiments at different $\sqrt{s}$ reproduce a similar pattern, but in 
different $x$ intervals.} since the various experiments probe different 
ranges of $x \simeq 
x_T = 2p_T/\sqrt{s}$.  On the other hand it has been shown \cite{HUSTON} that 
the discrepancy 
can be removed by a broadening of the transverse momenta of the initial state 
partons (due to multigluon emission) which increases with the energy $\sqrt{s}$.  
A similar effect has been quantified in Drell-Yan production, but here, so far, 
the broadening is accounted for phenomenologically.  Until the multigluon 
effects 
are calculated in QCD it is not possible to use the prompt photon data 
(especially 
those at higher energy) to pin down the gluon.  The most reliable determination 
comes from the lower energy $pp \rightarrow \gamma X$ data of the WA70 
collaboration, 
where the broadening is much less\footnote{Also $pp \rightarrow \gamma X$ has 
the advantage that the dominant LO subprocess is $gq \rightarrow \gamma q$, 
unlike $p\overline{p} \rightarrow \gamma X$ where $q\overline{q} \rightarrow 
\gamma g$ gives a comparable contribution.}, but even here there is an ambiguity 
of some $\pm$ 25\% in the value of the gluon.

\begin{center}
\subsection{Jet production at Fermilab}
\end{center}

Dijet production in $p\overline{p}$ collisions can also, in principle, probe 
the small $x$ behaviour of the gluon \cite{PPDJ,EWNG}.  For example, if the 
two jets are produced with equal transverse momentum $p_T$ but both very forward 
with pseudorapidity $\eta \gg 1$ then $x_1 \sim 1$ and $x_2 \sim (2p_T/\sqrt{s}) 
\exp (- \eta) \ll 1$.  Detailed NLO calculations \cite{EWNG} show that at 
$\sqrt{s} = 1.8$ TeV the gluon can be probed in this way in the range $0.005 < 
x_g < 0.05$.  However, at present the systematic errors are too large to allow 
any definite conclusion to be drawn.

The single jet inclusive cross section for jet transverse energies $E_T \sim 
100$ GeV is dependent on the gluon via the $gg, gq, g\overline{q}$ initiated 
subprocesses.  The gluon $g (x, Q^2)$ is sampled at $x \sim 2E_T/\sqrt{s} \sim 
0.1$ (and $Q^2 \sim E_T^2$) for centrally produced jets.  However, the $E_T$ 
spectrum gives more information on the running of $\alpha_S (E_T^2)$ than on 
the gluon.  In fact 
if we take the information on the gluon from the scaling violations of $F_2$ at 
HERA, then the jet $E_T$ spectrum gives a sensitive measure of $\alpha_S$ 
\cite{EJA}.  The steeper the spectrum the larger the prediction for $\alpha_S$.  
There are indications from the medium $E_T$ CDF (and also from the preliminary 
D0) jet data that the observed spectra favour $\alpha_S (M_Z^2)$ in the region 
0.116 to 0.120 \cite{EJA,MRSR}.  Again the systematic error is the limiting 
factor.

\begin{center}
\subsection{Dijet production at HERA}
\end{center}

The observation of dijets in deep inelastic scattering at HERA offers, in some 
respects, similar possibilities to jet production at Fermilab.  Again within a 
single experiment it is possible to observe the running of $\alpha_S (k_T^2)$.  
At HERA the LO subprocesses are the QCD Compton process $\gamma q \rightarrow 
gq$, and, relevant for the gluon, the $\gamma g \rightarrow q\overline{q}$ 
fusion 
reaction.  The NLO contributions are known and the scheme dependence has just 
been quantified.  Indeed Mirkes and Zeppenfeld \cite{MZ} have presented to this 
Conference a detailed study of the jet algorithms and conclude that the cone or 
$k_T$ schemes are favoured and lead to less scale dependence than the other 
schemes.  The clean identification and kinematic measurement of the jets is the 
experimental challenge.

\begin{center}
\subsection{Inelastic $J/\psi$ photoproduction}
\end{center}

It has long been advocated that inelastic $J/\psi$ photoproduction at HERA 
may serve as a measure of the gluon --- see, for example, ref.\ \cite{ELJPSI} 
which considers the colour-singlet model \cite{BJ} for the process at LO 
accuracy.  Recently the NLO contributions have been calculated \cite{KZSZ,KRA}.  
A detailed study of the spectra in the high energy range at HERA shows that 
the perturbative calculation is not well-behaved in the limit $p_T 
\rightarrow 0$, 
where $p_T$ is the transverse momentum of the $J/\psi$.  No reliable prediction 
can be made in this singular boundary region without resummation of large 
logarithmic corrections caused by multigluon emission.  If the small $p_T$ 
region is excluded from the analysis, the NLO result accounts 
for the energy dependence of the cross section and for the overall 
normalization, 
see Fig.\ 8 \cite{KRA}.  However, since the average momentum fraction of the 
partons is shifted to larger values when excluding the small-$p_T$ region, 
the sensitivity of the prediction to the small-$x$ behaviour of the gluon 
distribution is not very distinctive.

\begin{figure}[htb]
\begin{center}
~ \epsfig{file=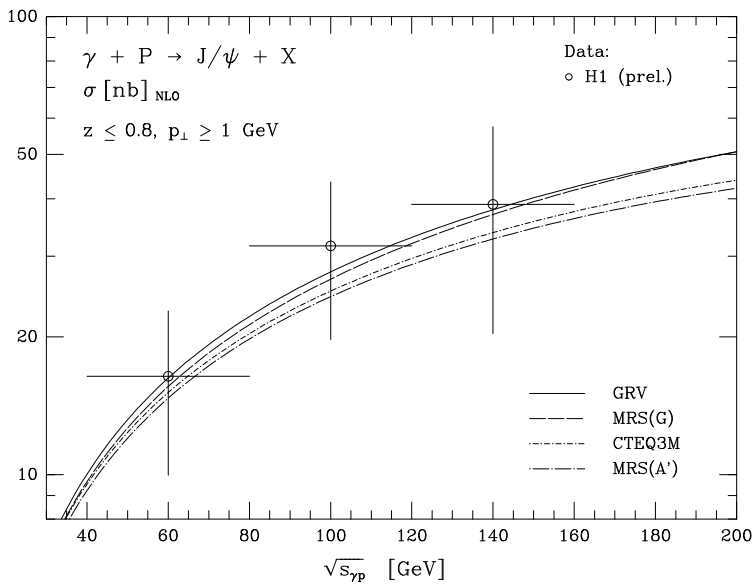,width=10cm,angle=-90}
\end{center}
\caption{\protect{\small Total cross section for inelastic $J/\psi$ 
photoproduction as a function 
of the photon-proton energy for different parametrizations of the parton 
distribution 
in the proton.  Experimental data from \protect{\cite{AID}}.  The figure is 
from \protect{\cite{KRA}}.}}
\label{fig8}
\end{figure}

\begin{center}
\subsection{Diffractive $J/\psi$ production at HERA}
\end{center}

Diffractive $J/\psi$ photoproduction appears to offer a more promising way to 
distinguish between the gluon distributions.  Since this is essentially an 
elastic process the cross section is a measure of the square of the gluon 
density.  To leading order the cross section is given by \cite{RYS,BFGMS}
\begin{equation}
\left . \frac{d \sigma}{d t} \: (\gamma^* p \rightarrow J/\psi p) \right |_0 \; 
= \; \frac{\Gamma_{ee} M_\psi^3 \pi^3}{48 \alpha} \: \frac{\alpha_S 
(\overline{Q}^2)^2}
{\overline{Q}^8} \; [xg (x, \overline{Q}^2) ]^{2}
\end{equation}
with $\overline{Q}^2 = \frac{1}{4} M_\psi^2$ and $x = M_\psi^2/W^2$, where $W$ 
is the $\gamma p$ c.m.\ energy.  In a recent study \cite{RRML}, corrections to 
this formula have been calculated and comparisons 
with HERA data made, see Fig.\ 9.  It was emphasized that the $W$ dependence, 
rather than the normalisation, was the more reliable discriminator between the 
gluons.  The power of the method is evident from Fig.\ 9, which appears to 
favour the MRS(A$^\prime$) gluon.  Further phenomenological studies of this 
process can be found in refs.\ \cite{PJS,H1JPSI}.

\begin{figure}[htb]
\begin{center}
~ \epsfig{file=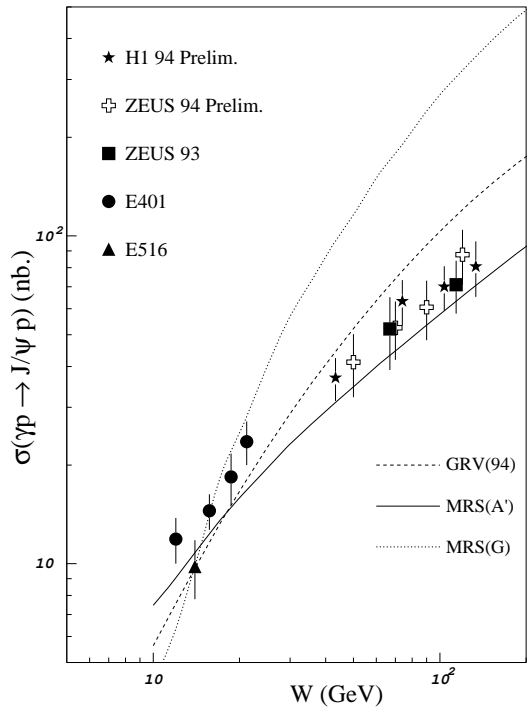,width=7cm}
\end{center}
\caption{\protect{\small The measurements of the cross section for diffractive 
$J/\psi$
photoproduction compared with the full perturbative QCD prediction obtained
from three 1994/5 sets of partons. The figure is taken from
\protect{\cite{RRML}}.}}
\label{fig9}
\end{figure}

\begin{center}
\section{The gluon at small $x$}
\end{center}

We have seen that the gluon is by far the dominant parton in the small $x$ 
regime.  Indeed in the perturbative region it drives the entire partonic 
structure of the proton via the $g \rightarrow gg$ and $g \rightarrow 
q\overline{q}$ transitions.

So far our description of the data, including the small $x$ HERA data down to 
$Q^2 = 1.5$ GeV$^2$, has been based on the DGLAP resummation of LO and NLO 
$\ln Q^2$ terms.  In fact the rise of $F_2$ with decreasing $x$ that is 
observed 
at HERA appears to be well described by the simplest approximation for the 
small 
$x$ behaviour of the gluon.  If in the DGLAP evolution for gluon, we take the 
splitting function equal to its small $x$ limit, $P_{gg} \simeq 
(3 \alpha_S/\pi)/x$, then it follows that
\begin{equation}
xg (x, Q^2) \: \sim \: xg (x, Q_0^2) \: \exp \left ( 2 \: \left [ \frac{36}{25} 
\: \ln \left ( \frac{t}{t_0} \right ) \: \ln \left (\frac{1}{x} \right ) \right 
]^{\frac{1}{2}} \right )
\label{eq:f}
\end{equation}
where $t \equiv \ln (Q^2/\Lambda^2)$, modulo slowly varying logarithms, provided 
the input $xg (x, Q_0^2)$ is not singular.  That is, in this double logarithm 
approximation, $xg$ increases faster than any power of $\ln (1/x)$, but slower 
than a power of $(1/x)$.  This behaviour feeds through into $F_2$ and gives an 
excellent description of the data, as emphasized by Ball and Forte \cite{BF}; 
the steepness of the rise in $F_2$ can be tuned to the data by adjusting the 
evolution length $Q^2/Q_0^2$.

Does the success of the DGLAP description of the HERA data indicate the 
dominance 
of the $\ln Q^2$ resummations and the absence of higher twists?  Such a 
conclusion 
would be premature.  At small $x$, $x \lapproxeq 10^{-3}$, we have, so far, only 
one type of data $(F_2^{ep})$ and there is freedom in the description, 
particularly 
as we have to supply the non-perturbative input at some scale $Q_0^2$.  
Clearly at 
sufficiently small $x$ the (NLO) DGLAP evolution will break down.  When 
$\alpha_S 
\ln 1/x \sim 1$ we have to also resum $\alpha_S \ln 1/x$ contributions 
(or, to be 
more precise, a whole series of $\ln^m Q^2 \ln^n 1/x$ terms).  At LO, 
the $(\alpha_S \ln 1/x)^n$ resummation is accomplished by the BFKL equation 
\cite{BFKL}.  In a physical gauge, the $(\alpha_S \ln 1/x)^n$ term corresponds 
to an $n$-rung effective ladder diagram (Fig.\ 2) in which the soft gluon 
emissions are strongly-ordered in longitudinal momenta, but in which the 
transverse momenta are no longer ordered.  Due to the latter fact we have to 
introduce the gluon distribution $f (x, k_T^2)$ unintegrated over $k_T^2$, and 
anticipate a diffusion in $\ln k_T^2$ as we proceed along the gluon chain.  The 
relation to the conventional gluon is given by
\begin{equation}
xg (x, Q^2) \; = \; \int^{Q^2} \: \frac{dk_T^2}{k_T^2} \: f (x, k_T^2).
\end{equation}
In principle, in the small $x$ domain the unintegrated distributions are the 
universal parton distributions which link process to process, via the $k_T$ 
factorization theorem \cite{KTFAC}.  An example of the theorem is given below 
in (\ref{eq:e}).

For fixed $\alpha_S$ the $x \rightarrow 0$ behaviour of the BFKL solution can 
be 
written in analytic form.  Keeping only essential factors it behaves as 
\begin{equation}
f (x, k_T^2) \: \sim \: x^{- \lambda_L} \: \exp \left ( \frac{- \ln^2 
(k_T^2/A^2)}
{B \ln (1/x)} \right )
\label{eq:d}
\end{equation}
where $\lambda_L$, the famous BFKL intercept, is given by $\lambda_L = 
(3 \alpha_S/\pi) 
4 \ln 2$.  Thus we have a $x^{- \lambda_L}$ power-like growth accompanied by 
a diffusion in $\ln k_T^2$.  If a physically reasonable prescription for the 
running of $\alpha_S$ is assumed then the BFKL equation may be solved 
numerically 
to yield \cite{AKMS} a form
\begin{equation}
f (x, k_T^2) \: \sim \: C (k_T^2) \: x^{- \lambda}
\end{equation}
where $\lambda \approx 0.5$ is less sensitive to the phenomenological treatment 
of the infrared region, than the normalisation $C$.  The BFKL prediction for 
the structure function $F_2$ is obtained using the $k_T$ factorization theorem 
\cite{KTFAC}
\begin{equation}
F_2 (x, Q^2) \; = \; \int_x^1 \: \frac{dx^\prime}{x^\prime} \: \int \: 
\frac{dk_T^2}{k_T^2} \: f (x^\prime, k_T^2) \: F_2^{\gamma g} \left ( 
\frac{x}{x^\prime}, k_T^2, Q^2 \right ),
\label{eq:e}
\end{equation}
where the integration is over the kinematic variables $(x^\prime, k_T^2)$ of 
the virtual gluon coupling to the quark box in Fig.\ 2.  $F_2^{\gamma g}$ is 
the off-shell gluon structure function which at LO is given by the quark box 
(and crossed-box) contributions to photon-gluon fusion.  The BFKL approach is 
also 
found to give a satisfactory description of $F_2$ for small $x$.  However, there are, 
at present, limitations to the \lq\lq prediction".  We comment on the 
ambiguities in the BFKL calculation of $F_2$ below.
\begin{itemize}
\item[(i)] Due to the diffusion of $f (x, k_T^2)$ in $\ln k_T^2$ there is a 
significant contribution from the infrared $k_T^2$ region which is beyond the 
scope of perturbative QCD and which has to be included using physically 
motivated 
phenomenological forms.  This leads to an uncertainty in the overall 
normalization 
of $F_2$, but much less in the $x$ dependence.  A physically reasonable 
treatment of 
the infrared region is found to give the experimental normalization.  In a sense 
this is equivalent to providing the non-perturbative input for DGLAP, but here 
the $x^{- \lambda}$ behaviour at small $x$ is prescribed.

\item[(ii)] The BFKL equation only resums the LO $\ln 1/x$ terms.  The NLO 
contributions are needed for a stable prediction.  Sub-leading effects, which 
to a large extent embrace both energy-momentum conservation and angular ordering, 
have been shown \cite{PJS,KMS} to significantly reduce the value of the exponent 
$\lambda$.

\item[(iii)] An underlying soft Pomeron contribution has to be included in the 
small $x$ 
region, determined by the extrapolation of the observed values of $F_2$ at large 
$x$.  Again this has the effect of reducing the value of $\lambda$ apparent in 
$F_2 \sim x^{- \lambda}$.

\item[(iv)] Shadowing corrections to the BFKL equation will eventually, as $x$ 
decreases, suppress the $x^{- \lambda}$ growth.  Although they have not yet been 
fully formulated, the evidence from the observed ratio of diffractive to 
non-diffractive deep-inelastic events, and from the persistent rise of $F_2$ at 
very low $Q^2 \approx 2$ GeV$^2$, indicates that shadowing effects are at most 
10\% in the HERA regime.

\item[(v)] We need further studies of a unified approach which incorporates, on 
a sound theoretical footing, both the BFKL and DGLAP resummations.
\end{itemize}

From the above discussion it is clear that there are many issues to be resolved.  We 
see that it will not be easy to quantify the importance of the $\ln 1/x$ 
BFKL-type contributions by using the effective $\lambda$ dependence of the 
measured values of $F_2$ at small $x$, that is $F_2 
\sim x^{- \lambda_{\rm eff}}$.

There has recently been much activity \cite{ANOM} based on expanding the 
anomalous dimensions $\gamma_{ij} (\alpha_S, \omega)$ in terms of $\alpha_S$ 
and the moment variable $\omega$.  For instance, for the gluon anomalous 
dimension, 
(LO) DGLAP resummation amounts to summing the terms
\begin{equation}
\gamma_{gg} \; = \; d_1 \: \frac{\alpha_S}{\omega} \; + \; d_2 \: \alpha_S \; 
+ \; d_3 \: \alpha_S \omega \; + \; \ldots ,
\label{eq:b}
\end{equation}
whereas BFKL resums a different subset of terms
\begin{equation}
\gamma_{gg} \; = \; b_1 \: \frac{\alpha_S}{\omega} \; + \; b_4 \: 
\frac{\alpha_S^4}{\omega^4} \; + \; \ldots
\label{eq:c}
\end{equation}
with the $\omega^{-n}$ term corresponding to a $x^{-1} \log^{n - 1} 1/x$ 
contribution to $P_{gg}$ (except for $n = 1$ where $P_{gg} \sim 1/x$, 
see (\ref{eq:a})).  Both the above expansions start with the same \lq\lq double 
logarithm" term
\begin{equation}
\gamma_{gg} \; = \; \int_0^1 \: dx \: x^\omega P_{gg} \; \approx \; \int_0^1 \: 
dx \: x^\omega \: \left ( \frac{d_1 \alpha_S}{x} \right ) \; = \; d_1 \: 
\frac{\alpha_S}{\omega} ,
\label{eq:a}
\end{equation}
with $b_1 = d_1$ which leads to the behaviour displayed in (\ref{eq:f}).  
Some of the first few coefficients of the BFKL expansion 
(\ref{eq:c}) vanish, namely $b_2 = b_3 = b_5 = 0$ \cite{JA}.  For this reason 
much of the rise of $F_2$ with decreasing $x$ is attributed to the LO expansion 
of $\gamma_{qg}$
\begin{equation}
\gamma_{qg} \; = \; \alpha_S \: \left ( c_1 \; + \; c_2 \: \frac{\alpha_S}
{\omega} 
\; + \; c_3 \: \frac{\alpha_S^2}{\omega^2} \; + \; \ldots \right ) ,
\end{equation}
which contributes to $F_2$ via $\partial F_2/\partial \ln Q^2 = P_{qg} \otimes 
g$.  All the coefficients $c_i$ are non-vanishing (unlike those for 
$\gamma_{gg}$), positive definite and large \cite{CH}.

In principle, this approach appears to offer the attractive possibility of 
quantifying the importance of $\ln 1/x$ effects by studying DGLAP-type 
evolution with anomalous dimensions (and coefficient functions) which 
incorporate 
the $(\alpha_S/\omega)^n$ terms.  However, it has been pointed out \cite{RSS} 
that such a procedure masks the true dependence on contributions from the 
infrared region.  Due to 
the diffusion in $\ln k_T^2$ at small $x$, an $(\alpha_S/\omega)^n$ 
contribution, 
which in DGLAP is assigned to the local point $Q^2$, actually samples the region 
of (the logarithm of) virtuality
\begin{equation}
\ln Q^2 \; \pm \; (\Delta (x))^n
\end{equation}
where $\Delta$ increases approximately as $(\ln 1/x)^{\frac{1}{2}}$.  Due to the large 
numerical coefficient, $B = (3 \alpha_S/\pi) 56 \zeta (3)$ (with $\zeta (3) = 
1.202$), in the diffusion term in (\ref{eq:d}), there is considerable implicit 
penetration into the infrared region.  For example it is found that the 
$(\alpha_S/\omega)^4$ term samples virtualities down to $Q^2/100$ \cite{RSS}.  
Thus it seems that there is no alternative but to work with the unintegrated 
gluon distribution and the $k_T$ factorization theorem and to study the effects 
of contributions from the infrared $k_T$ region explicitly.

The observable $F_2$ is too inclusive to show all the characteristics of the 
small $x$ properties of the unintegrated gluon $f (x, k_T^2)$.  In particular 
the $\ln k_T^2$ diffusion pattern is integrated over.  For this reason other 
observables which measure properties of the final state in deep inelastic 
scattering have been advocated as better indicators of $\ln 1/x$ resummation 
effects (see, for example, the reviews listed in ref.\ \cite{FIN}).

\begin{center}
\section{Conclusions}
\end{center}

The universal parton distributions of the proton are well determined by a wide 
range of data in the region $x \gapproxeq 0.02$.  The exception is the gluon, 
which although constrained, still has some residual ambiguity.  However, the 
scaling violations observed in the new, more precise HERA measurements of $F_2$ 
have pinned down the gluon in the small $x$ region ($x \sim 10^{-3}$).  One 
consequence is that the GRV model which gave such an excellent description down 
to $Q^2 \sim 1$ GeV$^2$, shows some systematic discrepancy.  The prediction is 
above the new HERA measurements at the lower values of $x$, see Fig.\ 6.  
Indeed the previous spread of possible gluon behaviour at small $x$ 
(represented by the MRS (A$^\prime$, G), CTEQ3 and GRV curves in Fig.\ 7) 
has been narrowed in global analyses \cite{MRSR} incorporating the new data to 
give gluons similar to that of the MRS(A$^\prime$) set.  Motivated by the 
success of the GRV approach, the latest global analysis \cite{MRSR} uses a 
lower input scale, $Q_0^2 = 1$ GeV$^2$.

The measurements of diffractive $J/\psi$ photoproduction at HERA were seen to 
also favour the gluon of the MRS(A$^\prime$) set of partons, see Fig.\ 9.  We 
briefly reviewed other ways in which the gluon may be measured.  Jet production 
at Fermilab and at HERA offer not only a constraint on the gluon, but also 
provide 
a sensitive measure of the running of $\alpha_S$.

Finally, we briefly discussed the perturbative QCD expectations for the 
behaviour of the 
gluon in the small $x$ region.  We emphasized the importance of resumming the 
$(\alpha_S \ln 1/x)^n$ terms when $x$ is sufficiently small so that $\alpha_S 
\ln 
1/x \sim 1$.  We highlighted the problems of incorporating these effects in the 
description of $F_2$ and the necessity to use the unintegrated gluon 
distribution together with the $k_T$ factorization theorem.  The dramatic 
improvement in the experimental measurements in the small $x$ domain serves as 
a challenge to provide a deeper theoretical understanding of this fascinating 
frontier of 
QCD.

\begin{center}
\section*{Acknowledgements}
\end{center}

It is a pleasure to thank Dick Roberts and James Stirling, and Jan Kwieci\'nski 
and Peter Sutton for most enjoyable research collaborations on the structure of 
the proton and Misha Ryskin for valuable discussions.  Also to thank our hosts 
in Krak\'ow for arranging an excellent Workshop.

\end{document}